\documentclass[10pt,conference]{IEEEtran}
\usepackage{cite}
\usepackage{amsmath,amssymb,amsfonts}
\usepackage{algorithmic}
\usepackage{graphicx}
\usepackage{textcomp}
\usepackage{xcolor}
\usepackage[hyphens]{url}
\usepackage{fancyhdr}
\usepackage{hyperref}
\usepackage{enumerate}

\newcommand{\hpcayear}{2027}

\title{Runtime Compilation for Quantum Circuits with PQF-Enabled $R_z$ Synthesis}

\newcommand\hpcaauthors{First Author$\dagger$ and Second Author$\ddagger$}
\newcommand\hpcaaffiliation{First Affiliation$\dagger$, Second Affiliation$\ddagger$}
\newcommand\hpcaemail{Email(s)}

\newcommand{\revision}{\textcolor{black}}

\newcommand{\protocol}{$\text{R}^2$-PQF }
\newcommand{\rp}{$Rx(\frac{\pi}{4})$}
\newcommand{\later}{\textcolor{red}}
\newcommand{\ignore}[1]{}

\author{
  \ifdefined\hpcacameraready
    \IEEEauthorblockN{
      \parbox{0.95\textwidth}{\centering
        \hpcaauthors{}
      }
    }
    \IEEEauthorblockA{
      \hpcaaffiliation{} \\
      \hpcaemail{}
    }
  \else
    \IEEEauthorblockN{
      \parbox{0.95\textwidth}{
        \centering
        Siddharth Dangwal\textsuperscript{1*},
        Christopher Kang\textsuperscript{1},
        Jad Habhab\textsuperscript{2},\\
        Ritvik Sharma\textsuperscript{3},
        Frederic T. Chong\textsuperscript{1},
        Gokul Subramanian Ravi\textsuperscript{2},
      }
    }
    \IEEEauthorblockA{
      \centering
      \textsuperscript{1}University of Chicago,
      \textsuperscript{2} University of Michigan,
      \textsuperscript{3} Stanford University
    }
  \fi
}

\fancypagestyle{camerareadyfirstpage}{%
  \fancyhead{}
  
  \fancyhead[C]{
    \ifdefined\aeopen
    \else
      \ifdefined\aereviewed
      \parbox[][12mm][t]{13.5cm}{\hpcayear{} IEEE International Symposium on High-Performance Computer Architecture (HPCA)}
      \else
      \ifdefined\aereproduced
      \parbox[][12mm][t]{13.5cm}{\hpcayear{} IEEE International Symposium on High-Performance Computer Architecture (HPCA)}
      \else
    \fi 
    \fi 
    \fi 
    \ifdefined\aeopen 
      \includegraphics[width=12mm,height=12mm]{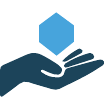}
    \fi 
    \ifdefined\aereviewed
      \includegraphics[width=12mm,height=12mm]{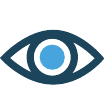}
    \fi 
    \ifdefined\aereproduced
      \includegraphics[width=12mm,height=12mm]{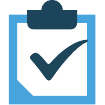}
    \fi
  }
  \fancyfoot[C]{}
}
\begin{document}
\maketitle

\begingroup
\renewcommand{\thefootnote}{\fnsymbol{footnote}}
\footnotetext[1]{Correspondence: \texttt{siddharthdangwal@uchicago.edu}}
\endgroup

\ifdefined\hpcacameraready 
  \thispagestyle{camerareadyfirstpage}
  \pagestyle{empty}
\else
  \thispagestyle{plain}
  \pagestyle{plain}
\fi

\newcommand{\hpcaheight}{0mm}
\ifdefined\eaopen
\renewcommand{\hpcaheight}{12mm}
\fi

\begin{abstract}
In Fault-Tolerant Quantum Computing (FTQC) using surface codes, reducing $T$-count is critical 
for efficient computation
due to the high cost of $T$-state distillation. \emph{Probabilistic Quantum Circuit with Fallback} (PQF) synthesis significantly lowers $T$-counts compared to standard protocols like Gridsynth, but introduces high circuit dynamism through extra gates inserted at runtime. This dynamism invalidates static compile-time routing, as these new $T$ gates require dynamic lattice surgery routing to establish non-intersecting ancilla paths to $T$-state factories while minimizing total execution time.

To address this, we introduce \textit{\protocol}, a \underline{R}untime \underline{R}outing protocol for \underline{PQF}-synthesized quantum circuits. \protocol dynamically routes gates layer-by-layer, computing paths during the execution of the previous logical timestep. To satisfy the strict computational window of $\sim \mathcal{O}(t_{lattice\_surgery} + t_{reaction})$ and non-intersectionality constraints, we implement a Wavefront Breadth-First Search (W-BFS) and synthesise it on a 500MHz FPGA and 2 GHz ASIC. For a clock cycle of 2GHz, we achieve a median logical time-step of $1.51 \mu s$ for a 300 qubit Ising Hamiltonian circuit, demonstrating that \protocol successfully resolves vast majority of paths (with rare exceptions) within the execution window and enables the $T$-count benefits of PQF without latency penalties. Compared to a baseline stalling-based solution (the only no-runtime-recomputation solution), \protocol        speeds up circuit execution by $1.34\times$ on average and up to $1.76 \times$.

\end{abstract}

\section{Introduction}\label{sec:introduction}

\begin{figure}
    \centering
    \includegraphics[width=\linewidth]{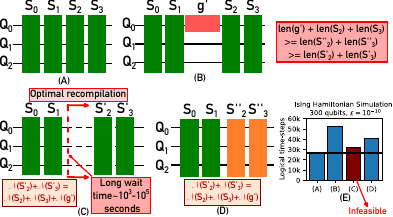}
    \caption{Scheduling a PQF circuit when synthesis fails at runtime.
\textbf{(A) Original Schedule (OS):} compile-time optimal schedule.
\textbf{(B) Stalling Schedule (SS):} synthesis failure inserts extra gates $g'$ at runtime. The only compile-time-viable response is to stall the OS until $g'$ completes, substantially increasing circuit duration.
\textbf{(C) Global Recompilation Schedule (GRS):} recompiling the pending gates ($S_2$, $S_3$) together with $g'$ minimizes duration, but optimal recompilation far exceeds the qubit coherence time and is infeasible.
\textbf{(D) \protocol Schedule (R2S):} \protocol recompiles $g'$ alongside the pending gates within the runtime budget. Not optimal, but feasible and shorter than stalling.
\textbf{(E)} The four schedules for a 300-qubit Hamiltonian Simulation circuit. \protocol gives the best feasible runtime schedule.
$len(S)$: duration of time step $S$. $\mathcal{N}(S)$: number of gates in time step $S$.}
    \label{fig:intro_fig}
\end{figure}

Quantum Error Correction (QEC) \cite{gottesman1997stabilizer, shor_ftqc} is essential for implementing flagship quantum algorithms such as Shor's, Grover's, and Hamiltonian Simulation \cite{shor1994algorithms, grover1996fast}.
The surface code is currently a leading candidate for QEC due to its high threshold and compatibility with nearest-neighbor connectivity \cite{Fowler_2012, fowler2018low}.
Various tools for compiling surface code circuits to future fault-tolerant quantum computers have been developed.
Generally, these tools take a logical quantum circuit and compile it to quantum hardware in the following steps: (1) Circuit synthesis: lowering the circuit to a fault-tolerant basis (for example, Clifford+$T$ operations), (2) Mapping: assigning each logical qubit and each $T$-magic state factory to a set of physical qubits, and finally (3) Routing: using lattice surgery to implement gates such as the logical CX and $T$ gates.

\textbf{Circuit Synthesis:} While surface codes can implement Clifford operations natively, non-Clifford operations must be synthesized into a Clifford+$T$ basis \cite{eastin2009restrictions}.
This synthesis generally uses static decomposition tools like Gridsynth \cite{ross2016optimal}, which produce fixed Clifford+$T$ decompositions of $R_z$ gates, resulting in a static circuit, i.e., one that does not change at runtime. \revision{The Clifford gates from these sequences can be eliminated using methods like TACO \cite{wang2026transpiler} resulting in a sequence of $T/T^{\dagger}/$\rp gates which costs much lesser than the original Clifford + $T$ sequence. All three gates - $T/T^{\dagger}/$\rp are equivalent from an implementation perspective, and hence we shall refer to them collectively as $T$ gates and the combined sequence as simply a $T$ sequence (unless explicitly stated otherwise).}

The compiled circuit's static nature allows layout, routing, and scheduling to be highly optimised before execution. The use of static schemes like Gridsynth comes at a cost: longer $T$ sequences, which inflate the $T$-count. Each $T$ gate requires a high-fidelity magic state produced by a Magic State Factory (MSF) \cite{litinski2019magic}, and these factories are extremely resource-intensive, with a large space-time overhead. Hence, there is a strong incentive to reduce the program's $T$-count.

\textbf{Dynamic Execution for Efficient QEC Circuits:}
While current compilation stacks use static methods like Gridsynth, dynamic synthesis methods like \emph{Probabilistic Quantum Circuits with Fallback} (PQF) offer a powerful alternative \cite{bocharov2014efficient, kliuchnikov2023shorter}. PQF offers a significant reduction in $T$-count over Gridsynth, at the cost of increased circuit dynamism.

\begin{figure}
    \centering
    \includegraphics[width=0.8\linewidth]{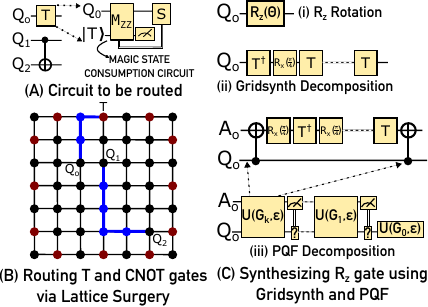}\caption{\revision{(A) Circuit to be routed. $T$ gate is implemented using \emph{Magic State Consumption} circuit which requires lattice surgery operations between $Q_o$ and Magic State. If the measurement output is $0$, a $T$ is implemented while an output of $1$ gives $T^{\dagger}$. For the \rp gate, we perform a $M_{ZX}$ lattice surgery operation, which has the same ancilla path and implementation cost as $M_{ZZ}$ (B) Routing paths for circuit (A)'s CX and $T$ gates (C) Gridsynth and PQF circuits to synthesize a $R_z$ rotation}}
    \label{fig:gs_pqf_and_routing}
\end{figure}

Historically, the architectural impact of the non-determinism inherent to FTQC has generally been managed using conservative fallback schemes. For instance, $T$ production failures are often countered by buffering many $T$ states, and a standard $T$-injection failure only requires an $S$-gate correction (Figure \ref{fig:gs_pqf_and_routing}(B)), which is implemented by stalling operations on the $S$-gate operand for one cycle.

PQF, however, introduces a drastically different scale of dynamism. PQF synthesizes each $R_z$ gate probabilistically, and the correctness of the synthesis relies on the measurement output of an ancilla qubit at runtime. If the synthesis attempt fails, a significant fallback circuit block (either another PQF trial or a lengthy Gridsynth sequence) must be dynamically inserted at runtime. Numerous operations in this block require lattice surgery paths. Further, the optimal routing decisions made at compile time are no longer valid.

\textbf{The Need for Runtime Recompilation:} The impact of PQF can be described generally as follows. We start with a schedule for a circuit that is optimized at compile time (Fig. \ref{fig:intro_fig}A), into which we want to insert a gate sequence $g'$ at runtime (Fig. \ref{fig:intro_fig}B).
We find that incorporating $g'$ into the circuit is non-trivial.

For example, on the one hand, we can handle this insertion by stalling the execution of the broader circuit until the dynamically inserted sequence $g'$ completes, as shown in Figure \ref{fig:intro_fig}B. By stalling, we ensure that the gates in $g'$ and those from the original circuit have no temporal overlap. Once all gates in $g'$ have been executed, we can continue using the original routing sequence. The gates in $g'$ are routed one per time step, using fixed, pre-computed paths. If we did not stall the original circuit, there would be potential for intersection between the lattice surgery paths of the original circuit's routes and the pre-computed paths of gates in $g'$. On the other hand, computing paths for $g'$'s gates at compile time in a way that is cognizant of the pre-existing lattice surgery routes is impossible. This is because the exact temporal overlap of gates can only be ascertained at runtime. \emph{Hence, stalling is the only solution to PQF routing that needs no runtime computation.} The stalling method, while effective, incurs massive latency penalties on execution. For a 300-qubit Ising Hamiltonian simulation circuit, the number of time steps almost doubles relative to the compile-time circuit (Figure \ref{fig:intro_fig}E). Fault-tolerant programs have time overheads from seconds to months \cite{beverland2022assessing, gidney2025factor}. Hence, a $2\times$ increase in time can be substantial.

Alternatively, one could attempt to fully recompile the remaining circuit along with the gates in $g'$ (Figure \ref{fig:intro_fig}C). Although a complete recompilation produces the shortest routing schedule, it can be notoriously slow. Attempting it at runtime is computationally prohibitive and often takes much longer than the qubit coherence time.

\textbf{Our Proposed Solution, \protocol:} To bridge the gap between the efficiency of PQF and the rigidity of static compilation, we propose \protocol, a lightweight, real-time recompilation step: \underline{R}untime \underline{R}outing protocol for \underline{PQF}-synthesized quantum circuits. Rather than starting from scratch, our approach builds on the existing static compilation and accounts for new circuit block insertions dynamically, but only incrementally, on the fly. \protocol follows the compile-time generated schedule until the gate sequence $g'$ is dynamically inserted at runtime. Instead of stalling or performing a costly global recompilation, \protocol recompiles the program layer by layer. It finds lattice surgery paths for gates in $g'$ and recomputes paths for logically concurrent gates from the original circuit to avoid intersections, increase gate throughput per time step, and minimize program duration. Crucially, \protocol completes the incremental routing path computation for the \revision{gates in time-step $i+1$ while the gates in time-step $i$ are executing on the quantum hardware} giving it a time budget of $\sim\mathcal{O}(t_{lattice\_surgery} + t_{reaction})$. For superconducting systems, this is on the order of microseconds.

Thus, \protocol processes dynamically inserted gate sequences at runtime. We visualize this in Figure \ref{fig:intro_fig}D, with the expected circuit depths resulting from different schemes in Figure \ref{fig:intro_fig}E.

\textbf{Key Contributions and Results.} We demonstrate that achieving the $T$-count benefits of PQF is practically feasible without incurring the latency penalties of traditional dynamic execution. Our specific contributions include:
\begin{itemize}
    \item \textbf{Identification of the PQF Routing Bottleneck:} We characterize the specific architectural challenges introduced by heavy runtime circuit dynamism, contrasting them with traditional quantum non-determinism.
    \item \textbf{The \protocol Architecture:} We design a lightweight runtime router that resolves ancilla paths for lattice surgery at runtime without violating non-intersectionality constraints.
    
    \item \textbf{Hardware Implementation with FPGA/ASIC:} We implement \protocol's dynamic routing routine for both an ASIC and an FPGA using AMD's Vitis HLS. For the ASIC implementation, we target TSMC 16nm FinFET technology, synthesized with Synopsys Design Compiler for the N16ADFP standard cell library. We also synthesize the RTL onto a Xilinx XU47P FPGA with Vivado.
    \item \textbf{System Evaluation:} For a 2 GHz clock, we achieve a median logical time step of $1.51\,\mu s$ in a 300-qubit Ising Hamiltonian circuit. We demonstrate that \protocol easily executes within the $\sim\mathcal{O}(t_{lattice\_surgery} + t_{reaction})$ time window. \revision{Compared to a baseline stalling-based solution, \protocol speeds up circuit execution by $1.34\times$ on average and by up to $1.76 \times$. Compared to a baseline runtime solution, \protocol speeds up circuit execution by $1.25\times$ on average and up to $1.59 \times$.}
\end{itemize}

\section{Background}\label{sec:background}

We point the reader to \cite{nielsen2001quantum, ding2022quantum, aaronson2013quantum} for background on quantum computing and \cite{Terhal_2015, gottesman1997stabilizer} for QEC. 

\subsection{Surface codes and $R_z$ synthesis}\label{subsec:surface_codes_synthesis}
For surface codes implemented on 2D architectures, multi-qubit Clifford gates, such as CX and $T$ gates are implemented via lattice surgery \cite{litinski2018lattice, Horsman_2012}. This involves the control and target qubits interacting with each other via a path of ancillas. \revision{For example, in Figure \ref{fig:gs_pqf_and_routing}A, the $M_{ZZ}$ operation requires a lattice surgery path connecting the $Z$ edges of $Q_o$ and the $T$ state patch as shown in Figure \ref{fig:gs_pqf_and_routing}B. For correctness, no two lattice surgery paths may intersect. Path intersections lead to the corresponding ancilla qubits being part of two different gates, which corrupts the signal in both of them. More details on path conflicts can be found in \cite{hamada2026bounded, molavi2025dependency}.}

To implement a general $n$-qubit quantum circuit, we first express it in the Clifford+$R_z$ basis. Each $R_z$ gate is subsequently decomposed into a $T$ sequence. Various methods can be used to perform this synthesis \cite{kliuchnikov2023shorter}; the most common is Gridsynth \cite{ross2016optimal}. In Gridsynth, the sequence is applied directly to the qubit that is the target of the $R_z$ rotation (Figure \ref{fig:gs_pqf_and_routing}C-i). The average number of gates in the decomposition scales as $\#T_{Gridsynth} = 3.02\log_{2}(\frac{1}{\epsilon})+1.77$ for synthesis precision $\epsilon$ \cite{kliuchnikov2023shorter}. Once the entire circuit is represented in the Clifford+$T$ basis, single-qubit Cliffords are implemented at low cost, and $CX$s and $T$s are implemented via lattice surgery. For the $T$ gate, one needs to produce a $T$ state using \textit{Magic State Distillation} (MSD) \cite{litinski2019magic}, followed by its consumption using the circuit shown in Figure \ref{fig:gs_pqf_and_routing}A. \revision{A measurement output of $0/1$ results in a $T/T^{\dagger}$ gate. For a \rp gate, instead of an $M_{ZZ}$ measurement, we need to perform an $M_{ZX}$ measurement, which requires the same ancilla path as $M_{ZZ}$; however, it connects to the $X$ edge of the $T$ state patch. Hence, implementationally, $T/T^{\dagger}/$\rp are equivalent, further justifying collectively representing all three as ``$T$''. A $T$ state patch is produced by a $T$ state factory located at the periphery of the layout (Figure \ref{fig:gs_pqf_and_routing}B).
MSD incurs a significantly higher overhead than any other operation, motivating the need to reduce the number of $T$ gates in the circuit \cite{litinski2019magic}. Although a recent proposal called \textit{Magic State Cultivation} (MSC) \cite{gidney2024magic} can produce $T$ states at a much lower space overhead (and hence space-time volume), the fidelity of these states cannot be increased arbitrarily due to an exponentially scaling post-selection requirement, making them unsuitable for large applications demanding high precision~\cite{huggins2025fluid}.}

\ignore{\revision{While Clifford gates are much cheaper than $T$, they are not free. Their contribution to the total circuit cost is increasing over time as $T$ production methods improve \cite{wang2026transpiler}. Recent work TACO by Wang et. al. \cite{wang2026transpiler} proposes a way to eliminate all Clifford gates out of a $H, S, T, \ldots$ sequence while preserving its logical correctness. The resulting gates in the sequence are $T, T^{\dagger}$ and \rp gates which can all be implemented using the same circuit with different lattice surgery measurements as shown in Figure \later{make the figure}. \emph{All three gates need the same magic state and have the same implementation cost.} For the rest of the paper, we will represent the synthesized sequences using $T, T^{\dagger}, $ \rp $\ldots$ sequences instead of $H, S, T, \ldots$ since they lead to shorter sequences at no implementation cost. Further, any reference to \emph{$T$ costs} or \emph{$T$ routes} applies to all three gates since they are implementationally equivalent.}}

A competing basis set to Clifford+$T$ is that of \emph{Pauli Product Rotations} (PPRs). Popularized by Litinski in Ref.~\cite{litinski2019game}, these involve multi-qubit Pauli rotations of non-Clifford angles. All Clifford gates are commuted to the end of the program and absorbed into the measurements. \revision{Although PPRs result in fewer gates than Clifford+$T$ circuits, each individual operation has a much larger support, often spanning the entire qubit register, which serializes the operations and results in much deeper circuits, as reported in TACO ~\cite{wang2026transpiler}. The high degree of parallelism coupled with TACO's Clifford elimination techniques, results in circuits that are on average $4.4\times$ and up to $21.9\times$ faster than PPRs (for a Gridsynth-facilitated Clifford+$T$ synthesis).} A comparison of PPR and PQF-synthesized circuits is presented in Section \ref{subsec:ppr_comparison}.

\subsection{Surface code compilation}\label{subsec:sc_compilation}

\begin{figure}
    \centering
    \includegraphics[width=\linewidth]{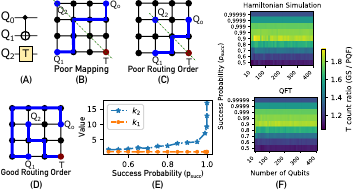}
    \caption{(A) Example circuit to be routed (B) Poor mapping prevents parallel execution of the CX and $T$ gates (C) Good mapping but poor routing order prevents parallel execution of the CX and $T$ gates (D) A good mapping and good routing order enables parallel execution of both gates (E) Variation of $k_1$ and $k_2$ parameters with probability of success ($p_{succ}$) of PQF, where $\#T_{PQF} = k_1\log_{2}(\frac{1}{\epsilon}) + k_2$ (F) Ratio of Gridsynth vs T gates across 10-400 qubits and $p_{succ} \in [0.5, 0.9999]$ for the a Ising Hamiltonian Simulation and QFT benchmarks. The synthesis precision $\epsilon = 10^{-5}$.}
    \label{fig:surface_code_compilation}
\end{figure}

As mentioned earlier, CX and $T$ gates are accomplished using non-intersecting ancilla paths. If any two paths intersect, we must serialize their execution, which increases circuit depth, execution time, and total error. Hence, a surface code compiler has two functions: (a) to map qubits such that the likelihood of intersections is minimal across all time steps, and (b) once a mapping has been fixed, to find ancilla paths that minimize serialization. Consider mapping and routing the circuit in Fig. \ref{fig:surface_code_compilation}a. The mapping in Fig. \ref{fig:surface_code_compilation}b is a poor one, since it serializes the CX and $T$ gates due to route conflicts. Fig. \ref{fig:surface_code_compilation}c shows a good mapping but a poor routing order: if we route the CX before the $T$ gate, we are forced to serialize the operations. Fig. \ref{fig:surface_code_compilation}d shows a good mapping together with a good routing, allowing the CX and $T$ gates to execute in parallel.
Numerous existing compilers address the surface code compilation problem \cite{molavi2025dependency, leblond2024realistic, beverland2022surface, hua2021autobraid} for static circuits. However, PQF produces dynamic circuits that differ drastically between compile time and runtime. Compiling these circuits remains an open problem to the best of our knowledge.

\subsection{Probabilistic Quantum Circuits with Fallback}\label{subsec:fallback_bg} 

The \textit{Probabilistic Quantum Circuits with Fallback} (PQF) algorithm is a probabilistic synthesis method that yields a much smaller number of $T$ gates than Gridsynth. Its decomposition depends not only on the precision parameter $\epsilon$, but also on a success probability parameter, $p_{succ}$. A PQF decomposition is implemented using the circuit shown in Figure \ref{fig:gs_pqf_and_routing}C-iii. Here, we require an extra algorithmic ancilla on which the gates of the decomposition are applied, all sandwiched between two CXs. This is followed by a measurement on the ancilla. A rotation of $R_z(\theta)$ ($R_z(-\theta)$) is implemented if the measurement succeeds (fails), with probability $p_{succ}$ ($1 - p_{succ}$). An incorrect rotation must be corrected by implementing $R_z(2\theta)$, which can be accomplished with another PQF round. Otherwise, we ``fall back'' to a deterministic strategy, Gridsynth in this case, and terminate the synthesis. The overall PQF schematic is shown in Figure \ref{fig:gs_pqf_and_routing}C-iii, where we perform $k$ rounds of PQF before falling back to Gridsynth at the end.
The average number of $T$ gates in PQF scales as $\#T_{PQF} = k_1\log_{2}(\frac{1}{\epsilon})+k_2$, where $k_1$ and $k_2$ are functions of the user-tunable parameter $p_{succ}$. For $p_{succ} = 0.99$, these values are $k_1=1.03$ and $k_2=5.75$, a $\sim3\times$ improvement in the leading coefficient relative to Gridsynth \cite{kliuchnikov2023shorter}. The variation of $k_1$ and $k_2$ with $p_{succ}$ is shown in Figure \ref{fig:surface_code_compilation}e. While $k_2$ increases rapidly as $p_{succ}$ grows, $k_1$ hovers close to 1 over the entire range of $p_{succ}$.
However, this estimate does not account for the Gridsynth gates inserted when synthesis fails. Figure \ref{fig:surface_code_compilation}f shows the ratio of $T$ gates in a circuit synthesized using Gridsynth versus PQF (accounting for synthesis failure) for Quantum Fourier Transform (\texttt{qft}) \cite{shor1994algorithms} and Hamiltonian Simulation (\texttt{hamsim}) \cite{lloyd1996universal} circuits at precision $\epsilon = 10^{-5}$. Across both benchmarks, the mean and maximum reductions in $T$ count are $1.49\times$ and $1.92\times$ respectively.
This ratio is highest in the range $p_{succ} \in [0.9, 0.99]$. A lower value of $p_{succ}$ ensures a smaller $T$ count in the PQF decomposition (smaller $k_1$ and $k_2$); however, the probability of synthesis failure, and hence of a subsequent insertion of Gridsynth gates, increases. A higher $p_{succ}$ results in fewer failures, but the original PQF synthesis has more $T$ gates (higher $k_1$ and $k_2$). The optimal point within this range depends on the benchmark and the precision, although exceptions may exist.


\begin{figure*}
    \centering
    \includegraphics[width=0.95\linewidth]{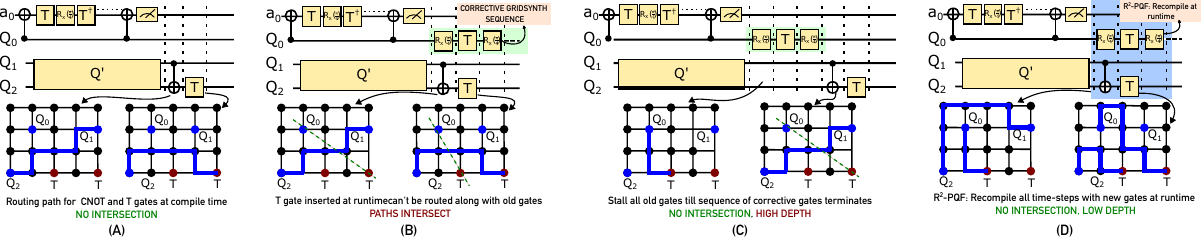}
    \caption{PQF failure and handling insertion of Gridsynth gate sequence at runtime (A) Before PQF failure (B) Insertion of Gridsynth gate sequence causes routing path conflicts (C) Stalling the original circuit's gates till Gridsynth sequence completes resolves path conflicts (\emph{Stalling Schedule}). (D) \protocol recompiles gates one time-step at a time resolving conflicts between runtime inserted Gridsynth gates and the original gates.  (\emph{\protocol Schedule}). }\label{fig:runtime_path_conflicts}
\end{figure*}

\begin{figure}
    \centering
    \includegraphics[width=0.9\linewidth]{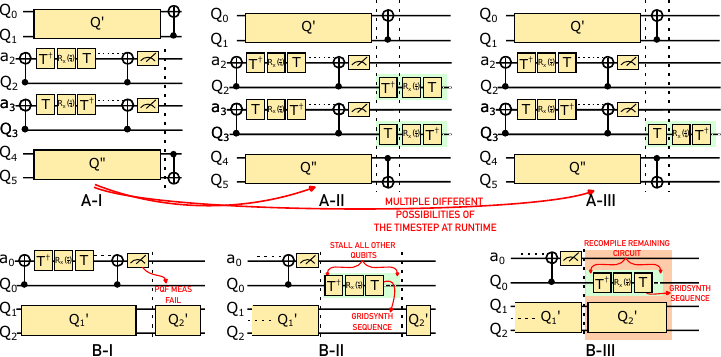}
    \caption{\textbf{A-I, II, III:} Different possibilities of the timestep at runtime based on which PQF measurements fail \textbf{B-I:} Original Schedule \textbf{B-II: }Stalling Schedule: Stall original circuit's remaining gates ($Q_2'$) till Gridsynth sequence completes execution \textbf{B-III:} Global Recompilation Schedule: Perform global recompilation of Gridsynth gates along with $Q_2'$ to obtain minimum depth schedule (infeasible at runtime).}
    \label{fig:r2_pqf_motivation}
\end{figure}


\section{Compiling PQF Circuits}\label{sec:motivation}


\subsection{Changes in PQF circuits at runtime}\label{subsec:changes_at_runtime}

As mentioned in Section \ref{subsec:sc_compilation}, the goal of a compiler is to minimize program duration. A shorter duration means a better compiler. The decisions made by a compiler depend on the input circuit's gate ordering. Since PQF circuits change at runtime, the original compiled schedule, which we call the \textit{Original-Schedule} (OS), does not account for the extra runtime-inserted gates.

A corrective Gridsynth sequence contains $T/T^{\dagger}/$\rp gates that can cause unforeseen path conflicts, which serialize computation, and subsequently increase program duration. For example, in Figure \ref{fig:runtime_path_conflicts}A, we implement an $R_z(\theta)$ operation on $Q_0$ using PQF, while qubits $Q_1$ and $Q_2$ undergo operations which we collectively represent as $Q'$. Right when the PQF sequence terminates, there is a CX between qubits $Q_1$ and $Q_2$, which is implemented using the routing path shown in blue. This path is trivially optimal since no other gates exist in this time step ($0$ intersections).

Figure \ref{fig:runtime_path_conflicts}B shows the resulting circuit if the PQF measurement fails at runtime. This necessitates the insertion of a corrective sequence of Gridsynth-synthesised gates on qubit $Q_0$ (highlighted in green) The execution of the first \rp gate in this sequence requires a lattice surgery path between $Q_0$ and a $T$ state factory at the periphery of the layout. One cannot create this path without intersecting with the pre-existing CX path also scheduled for the same time step. This forces serialization of computation. A similar conflict arises between the two concurrent $T$ gates in Figure \ref{fig:runtime_path_conflicts}B. 

\subsection{Handling runtime changes to PQF circuits}\label{sec:previous-work}
In the general case, it is infeasible to pre-compute the paths for runtime-inserted gates, prior to program execution. 
This is attributed to the stochastic nature of PQF synthesis, due to which one can't determine the gates that align in the same time-step at runtime prior to program execution.

For example, in Figure \ref{fig:r2_pqf_motivation}AI-III the time-step sandwiched between the two dashed lines can look quite different based on which PQF measurements fail, resulting in different routing paths. A circuit with $N_{rz}$ $R_z$ gates has $2^{N_{rz}}$ possible runtime structures, each corresponding to a different routing schedule. 
\ignore{For example, a 400-qubit trotterized Ising Hamiltonian simulation circuit, simulated for 3 time-steps, has 6809 $R_z$ gates \cite{Supermarq}, resulting in $2^{6809}$ different routing schedules!}


In this section, we will describe two solutions for handling PQF dynamism in circuits. These techniques, which are based on stalling and recompilation respectively, result in the \emph{Stalling-Schedule} (SS) and the \emph{Global Recompilation-Schedule} (GRS) (Figures \ref{fig:r2_pqf_motivation}B-II, III). SS is the only compile-time solution to PQF dynamism. Its duration acts as an upperbound for \protocol. GRS is an ideal lowerbound for the duration of \protocol. Generating GRS incurs a large time overhead, often orders of magnitude longer than program coherence time (Figure \ref{fig:intro_fig}C) which makes it practically infeasible. However, it is still useful for gauging the performance of \protocol.

\subsubsection{Upperbound: Stalling-Schedule}\label{subsec:stalling_sched}

One way to execute runtime-inserted $T$ gates is to pre-compute the path to the nearest $T$ state factory for each qubit in the circuit. 
At runtime, if a PQF measurement fails, we stall all operations except the runtime-inserted gates. Since only a single gate is executed per time step, there is no scope for path conflicts and any pre-computed routing path becomes viable. This requires no additional computation at runtime. The resulting schedule is the Stalling-Schedule (SS). Figures \ref{fig:runtime_path_conflicts}C and \ref{fig:r2_pqf_motivation}B-II show the SS. We follow OS before PQF measurement fails. Next, we stall the unexecuted gates in OS (The $CX$ and $T$ gates on qubits $Q_1$ and $Q_2$ in Figure \ref{fig:runtime_path_conflicts}C and the $Q_2'$ block in Figure \ref{fig:r2_pqf_motivation}B-II), till we finish routing all gates in the corrective sequence - one gate per time step. Once the Gridsynth sequence completes, we revert to the OS.

\begin{figure}
    \centering
    \includegraphics[width=\linewidth]{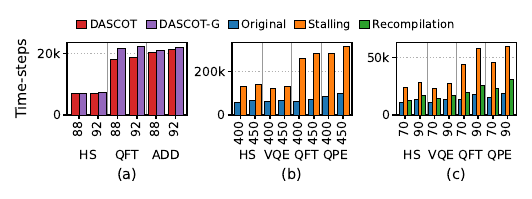}
    \caption{\textbf{(A) DASCOT vs DASCOT-G:} DASCOT-G gives comparable routing depths to DASCOT at substantially lower compilation time overheads. We show data for Hamiltonian Simulation (HS), Adder (ADD) and Quantum Fourier Transform (QFT) \textbf{(B) OS vs SS:} \emph{Stalling Schedule} (SS) is on average $2.78\times$ and up to $4.11 \times$ times longer than the \emph{Original Schedule} (OS). \textbf{(C) GRS vs SS: } \emph{Global Recompilation Schedule} (GRS) is on average $1.69\times$ and up to $2.19\times$ shorter than SS. See Section \ref{sec:motivation} for details.} 
\label{fig:dascot_g_and_baselines}
\end{figure}

\textbf{Increase in runtime:}
The forced serialization in SS results in a substantial increase in program duration, which we treat as an upper bound for the duration of the schedule generated by \protocol. Figure \ref{fig:dascot_g_and_baselines}B compares the duration of the SS with that of the OS for four benchmarks with 400 and 450 qubits. We observe that the SS is on average $2.78\times$ and at most $4.11\times$ longer than the OS.

While inefficient, the SS is a viable way to route PQF-synthesized circuits. It is also the only strategy that doesn't require any runtime computation. The moment we try to route a runtime-inserted gate in parallel with an unexecuted gate from the OS, we need to check for path conflicts, which subsequently requires runtime computation.

\subsubsection{Lowerbound: Global Recompilation-Schedule}\label{subsec:recompilation_sched} To obtain the shortest possible schedule after adding Gridsynth gates to the circuit, one needs to perform a global recompilation of the entire remaining circuit.  

When we perform a global recompilation, the compiler uses lookahead-based strategies and global context to optimize the routing order among commuting gates \cite{molavi2025dependency} and obtain the shortest possible sequence. We call the resulting schedule the Global Recompilation-Schedule (GRS). 

Figure \ref{fig:r2_pqf_motivation}B-III shows the GRS. We proceed with the routing paths in OS till we encounter a PQF measurement failure. We take all remaining gates in the circuit ($Q_{2}'$ along with the corrective Gridsynth sequence), and pass them through a compiler to obtain the schedule for the remaining circuit. We follow the routing paths in this recompiled schedule until we encounter another PQF measurement failure, at which point we recompile yet again. The amalgamation of all routing paths used in the circuit constitutes the GRS.

Figure \ref{fig:dascot_g_and_baselines}C shows the duration of GRS compared to that of OS and SS for circuits of 70 and 90 qubits across four benchmarks. Compared to OS, the GRS is on average $1.42\times$ and up to $1.78\times$ longer. Compared to SS, the GRS is on average $1.69\times$ and up to $2.19\times$ shorter.
Note that computing the GRS at runtime is often practically infeasible. Recompiling the entire program using high-performance software compilers takes multiple minutes to days (depending on circuit size). In this time, the circuit will stall and accumulate memory errors, which are often sufficient to corrupt the entire computation. However, the duration of GRS serves as an idealized lower bound for the schedule generated by \protocol.

\textbf{High compilation overhead}
For this paper, the GRS is generated using a greedy version of the state-of-the-art DASCOT compiler \cite{molavi2025dependency}, which we call ``DASCOT-G''. DASCOT uses a simulated annealing solver to optimize the number of time steps in the schedule. The compilation time is impractically long, even for static circuits. For example, DASCOT takes 2086 seconds to generate a 7162 time-step schedule for a 92-qubit Trotterized Ising Hamiltonian Simulation benchmark evaluated for one Trotter Step. Note that this is a static circuit which requires one-time compilation before execution. A dynamic version of the same circuit synthesized using PQF, with $p_{succ}=0.9$, results in $174$ failed $R_z$ syntheses on average (total 1738 $R_z$ gates), resulting in a total compilation overhead (and hence stalling time) of $174 \times 2086$s $= 100.82$ hours.

To make the evaluation loop more efficient, we use the greedy version - DASCOT-G, which reports comparable mapping and routing performance to DASCOT at a much lower time cost. For example, DASCOT-G compiles the 92-qubit Ising Hamiltonian Simulation circuit in 123 seconds ($17\times$ speedup) and generates a schedule with 7226 time steps ($0.9\%$ performance drop). Figure \ref{fig:dascot_g_and_baselines} (A) shows the compilation performance of DASCOT-G against DASCOT \cite{molavi2025dependency}. 

While DASCOT-G is much faster than DASCOT, it is still infeasible to use it to recompile the circuit at runtime. The runtime recompilation window is many orders of magnitude shorter than the 123 seconds taken by DASCOT-G as we will see in Section \ref{subsubsec:runtime_path_determination}. Waiting for such a long time results in long circuit stalls and subsequent loss in fidelity. \ignore{For example, Table 1 in Ref \cite{beverland2022assessing} assesses a $99.9\%$ algorithmic accuracy for quantum dynamics circuits. For a 7226 time-step schedule, this can be guaranteed by a surface code distance of $d \sim 15$ resulting in a program execution time of $\sim 0.3$ seconds. However, if we recompile at runtime using DASCOT-G, the program will stall for a total of $5.94$ hours waiting for new compilation schedules. Not only is this a $7.1\times10^4$ times increase in program duration, it is also more than enough to completely decay qubit coherence and turn the output into noise.} 

\subsection{Short schedule at low compilation overhead: Recompilation using \protocol}

The two prior solutions - SS and GRS highlight important challenges in routing dynamic PQF circuits. The former is practically feasible but produces a long depth schedule. The latter produces a compact schedule, but requires extensive runtime computation - often long enough for qubit decoherence \cite{beverland2022assessing}, making it impractical. We need a solution that is both \emph{practically implementable} and \emph{reduces the schedule's depth}. 

The only way to reduce the schedule's depth relative to SS is to route the runtime-inserted gates in parallel with the circuit's pre-existing gates. There are two requisites for this: 
\begin{enumerate}[a.]
    \item \emph{Path determination at runtime:} To maximize parallelism, the new route must be computed at runtime as discussed in Section \ref{sec:previous-work} since it needs to be cognizant of concurrent gates' routes.
    \item \emph{Non-intersection with existing routes:} There needs to be a valid, vacant ancilla route which can be used to implement the new gate without intersecting existing routes.
\end{enumerate}

We discuss these requisites using Figure \ref{fig:runtime_path_conflicts}

\subsubsection{Path determination at runtime}\label{subsubsec:runtime_path_determination} In Figure \ref{fig:runtime_path_conflicts}B, executing the first $T$ gate in the Gridsynth sequence requires a path between $Q_0$ and a $T$ factory. Unlike gates in OS, for which routing paths are known at compile time, the path for the new $T$ gate needs to be determined \textit{at runtime}. We discussed in Section \ref{sec:previous-work} and Figures \ref{fig:r2_pqf_motivation}AI-III that it is impossible to know the pre-existing gates that a runtime-inserted gate will concur with before circuit execution due to the probabilistic nature of PQF. Since concurrent routes need to be cognizant of each other to avoid intersections, we can't generate the routing path for a runtime-inserted gate at compile time (except in the case of the Stalling Schedule, because we eliminate concurrency by forcing the execution of one gate at a time). 

The process of computing the routing path at runtime needs to be fast. \revision{To avoid any program stalls due to compilation overheads, the compiler must generate a gate's routing paths before it is sent to the QPU. This is enforced by compiling in a sliding-windows fashion -- the routes for gates in time-step $i+1$ are computed while gates in time-step $i$ are executing. Hence, the compiler's time-budget is equal to the execution time of the previous layer. This is a small window - the exact value of which depends on the preceding operation, code-distance, measurement cycle time and reaction time.}

\revision{The time taken to implement $T/T^{\dagger}/$\rp gates is the same -- \underline{$t_{lattice\_surgery} + t_{reaction}$.} $t_{lattice\_surgery}$ is the time taken to implement a lattice-surgery operation. For a distance $d$ surface code, $t_{lattice\_surgery} = time(d \text{ measurement rounds})$ \cite{litinski2019game}. On superconducting hardware, a single measurement round is $\sim 1\mu s$ long \cite{IBMQE, acharya2024quantum}. This value is also used extensively in quantum architecture literature \cite{maurya2023scaling, maurya2024understanding, khalid2025impacts}. For neutral atom and trapped ion systems, measurement time is of the order of mili-seconds \cite{bluvstein2024logical, manovitz2022trapped}. For this paper, we will focus on the stricter time-constraint of $1 \mu s$. Hence, $time(d \text{ measurement rounds}) = d \mu s$. A discussion for neutral atoms and trapped-ions is in Section \ref{subsec:hardware_latency}.} 

\revision{$t_{reaction}$ or the ``reaction time'' of the computation is the time taken to process the results of a syndrome extraction cycle classically and communicate the action conditioned on this result to the quantum processor. This extra overhead occurs for all $T/T^{\dagger}/$\rp gates, since we can't commute a Pauli correction past them and apply it in software. Different references assume different heuristic values of $t_{reaction}$, however, the most commonly used on in resource-estimate papers is $10\mu s$ \cite{huggins2025fluid, gidney2025factor, litinski2022active, lee2021even}. Ref \cite{khalid2025impacts} performs an extensive modeling of reaction-time delays and concludes that a more realistic number is $\sim 100\mu s$. Overall, $t_{reaction}$ modeling is an active area of research with no conclusive, validated number. A larger value of $t_{reaction}$ results in a larger time budget for runtime recompilation. However, we choose the conservative value of $t_{reaction} = 10 \mu s$.}

\revision{Consequently, for a distance $d$ surface code, $t_{lattice\_surgery} + t_{reaction}=(d+10) \mu s$. For code-distance $d > 10$, we have a time-budget of $> 20 \mu s$. For utility scale circuits, the regime covered in this paper, one typically requires much larger code distances. For example, Microsoft's resource estimate paper uses $d = 21, 27, 33$ for quantum dynamics, factoring, quantum chemistry respectively \cite{beverland2022assessing}. Overall, the time-budget is highly variable, and $20 \mu s$ represents a \emph{lower-bound} over all feasible values of code distance, measurement time and reaction time for $T/T^{\dagger}/$\rp sequences. In practice, the time-budget would often be much larger, the implications of which are discussed in Section \ref{subsec:hardware_latency}.}

\begin{figure*}
    \centering
    \includegraphics[width=0.95\linewidth]{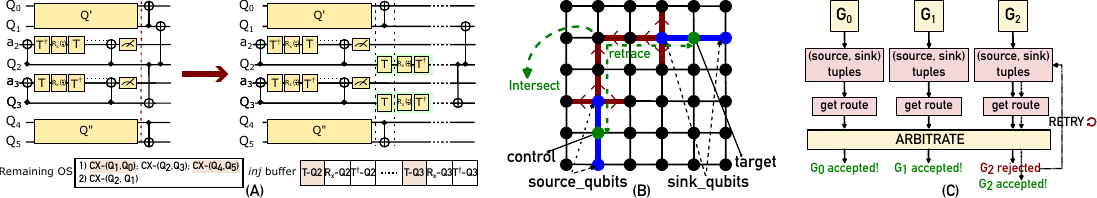}
    \caption{\textbf{(A) PQF measurement fails for $Q_2$ and $Q_3$: }All gates after PQF measurement failure (after the red dotted line) are part of the Remaining OS. Runtime inserted gates go into the \emph{inj} buffer. We sweep the Remaining OS and \emph{inj} buffer to determine gates with no impending data dependencies. This set is called the \emph{sched\_frontier} (highlited in light pink). \textbf{(B) Wavefront BFS:} Vertical (horizontal) neighbors of the control (target) qubit are the source (sink) qubits. A wavefront grows from both the source and sink qubits till it intersects (red arrows). The path from the intersection point to both source and sink qubits is retraced (dotted green). This is the gate's routing path \textbf{(C) Gate Arbitration:} We compute candidate routes for all gates in the \emph{sched\_frontier} in a given timestep. The computed routes are arbitrated for intersections. Non-intersecting paths are accepted and deleted from the \emph{sched\_frontier}. The intersecting paths are rejected and recomputed. If a path still intersects the accepted paths, the corresponding gate fails to get routed in the current time-step. It stays in the \emph{sched\_frontier} and its routing is re-tried in the following time-step.}
    \label{fig:r2pqf_design_1}
\end{figure*}

\subsubsection{Unresolvable conflicts with existing routing paths}\label{subsubsec:unresolvable_conflicts} Continuing with the CX path in OS (Figure \ref{fig:runtime_path_conflicts}A) leads to another issue. Assuming we could compute the path for the runtime-inserted $T$ gate within the $10 \mu s$ budget, to route both gates in parallel, there needs to exist a valid, vacant ancilla route between $Q_0$ and the $T$ factory, which does not intersect with the CX's path. As we see in Figure \ref{fig:runtime_path_conflicts}B, there is no such route here, forcing the compiler to serialize the CX and $T$ gates and increasing program duration. Such instances are ubiquitous across large programs (also see the two $T$ gates in the same time-step in Figure \ref{fig:runtime_path_conflicts}B).  

However, if the compiler were cognizant of the presence of the $T$ gate, it could route the CX gate differently and ensure that both the $T$ and CX gates get routed in parallel. This is shown in Figure \ref{fig:runtime_path_conflicts}D. Since the presence/absence of this $T$ gate is determined only at runtime, the compiler needs to not only find the path of the runtime-inserted gates, but also \textit{recompute} paths for gates in OS to enable the routing of the new gates in parallel with the old ones - all at runtime.

\subsection{\protocol for solving runtime routing}

Given the aforementioned requisites, the central question is: \textit{can we achieve a shorter schedule than SS by performing runtime computation that finishes within a feasible time budget?} Our work \protocol answers affirmatively, tackling the two issues in runtime routing: determining new routes and handling unresolvable conflicts with existing paths within the target time constraint. Next, we describe the design of \protocol. 

\section{\protocol Design}\label{sec:tech-core}

The \protocol algorithm starts with the OS, and modifies it at runtime to obtain the \textit{\protocol-Schedule} (R2S). The algorithm can be split into three cases: 

\subsection{Case 1: No failed PQF measurements yet}\label{subsec:no_failed_meas}
Before any PQF failure, we follow the OS since the circuit structure remains the same as the original circuit at compile time. Thus, using  OS is optimal.

\subsection{Case 2: PQF failure detected}\label{subsec:failure_detected}
When PQF fails, a stream of gates is injected into the circuit at runtime. The primary challenge is to route these gates parallely with the pending gates from OS to the greatest possible extent. 
To achieve this, we perform a per-timestep recompilation 
at runtime. The steps are:  

\subsubsection{Step 1 - Add pending gates to a buffer} All runtime-inserted gates are added to a buffer (called $inj$ buffer) while maintaining data dependencies among themselves (Fig \ref{fig:r2pqf_design_1}A).

\subsubsection{Step 2 - Determine all gates at the execution frontier} From the $inj$ buffer, and the set of pending gates in OS, we obtain all the gates that do not have any pending data dependencies, and act on unoccupied qubits. They form the program's execution frontier - a.k.a $sched\_frontier$ (Fig \ref{fig:r2pqf_design_1}A).

\subsubsection{Step 3 - Choose source and target candidates for each gate}
While transversal gates in $sched\_frontier$ can be implemented in-place, to execute CX and $T$ gates, we need to compute their routing paths.

In our design, we assume fixed, non-rotating surface code patches with vertical $Z$ edges and horizontal $X$ edges. \revision{Based on the required lattice surgery measurement, we choose vertical/horizontal starting and ending points of the path. For example, a CX gate requires a $ZZ$ measurement followed by an $XX$ measurement, a valid lattice surgery path starts at a vacant vertical neighbor of the control qubit and terminates at a vacant horizontal neighbor of the target qubit.} This design choice is similar to the DASCOT compiler \cite{molavi2025dependency}. Note that one can assume that the surface codes rotate with a single change in design - the routing path can start and terminate at any unoccupied neighbor of the control and target qubits, respectively.
We refer to the vacant vertical neighbors of the control qubit as the $source\_qubits$ and the vacant horizontal neighbors of the target qubits as the $sink\_qubits$. Generally, a CX gate has 2 $source\_qubits$ and 2 $sink\_qubits$. The number reduces if the control/target qubit is at the boundary of the routing grid or next to an occupied qubit. A $T$ gate has 2 $source\_qubits$ too; however, the number of $sink\_qubits$ depends on the number of MSFs in the layout.
We cap the $sink\_qubits$ to be the 10 closest horizontal neighbors of the MSFs. See Fig \ref{fig:r2pqf_design_1}B for the source and sink qubits.

\ignore{For the layout shown in Figure \ref{fig:gs_pqf_and_routing}B, all the vacant horizontal neighbors of the magic states (colored red) are valid $sink\_qubits$.
If a layout has a large number of MSFs, we cap the number of $sink\_qubits$ to 10 for the 10 closest valid patches.} 

\subsubsection{Step 4 - Compute routes for gates}\label{subsubsec:wbfs}For each gate, once $source\_qubits$ and $sink\_qubits$ have been defined, we iterate through all (source, sink) tuples in increasing order of their Manhattan distance. Ties are broken randomly. For each (source, sink) tuple, we use a Wavefront Breadth-First Search (W-BFS) to compute the path between them, similar to the hardware implementations of matching decoders \cite{das2022afs, NISQ+}. The W-BFS algorithm is especially suited to an FPGA/ASIC implementation, which is critical to meet the strict timing constraints required for a runtime router.

In the W-BFS algorithm, in each classical clock cycle, a node sends signals to all its vacant neighbors, a maximum of 4 signals per cycle. This results in exponentially growing source and sink wavefronts. The algorithm terminates once the source and sink wavefronts intersect. As soon as a valid intersection is detected, we terminate wavefront expansion and backtrack from the intersection point to obtain the routing path (Fig \ref{fig:r2pqf_design_1}B). The iteration through (source, sink) pairs is stopped.

\subsubsection{Step 5 - Arbitration among routes} Once a potential route for each gate in the current time-step has been obtained, we iterate through them one at a time, accepting only those that don't intersect any previously accepted routes. The iteration order is critical since a gate's acceptance/rejection depends on all the routes accepted before it. It needs to be one that \emph{maximizes gate throughput}. The following rules decide the iteration order in decreasing order of priority:
\begin{enumerate}
    \item \textbf{Number of (source, sink) tuples: } Routes corresponding to gates with \emph{fewer} (source, sink) candidates are prioritized since they have fewer routing path options. More (source, sink) candidates implies a higher likelihood of routing path discovery even when the layout has been populated by existing routes. \ignore{Hence, lowering their routing priority increases overall gate throughput.} 
    \item \textbf{Minimum Manhattan distance across (source, sink) tuples:} Routes corresponding to gates with a \emph{higher Manhattan distance lowerbound} are prioritized. A larger Manhattan distance lowerbound implies that the control and target qubits are farther away and require more vacant ancillae for routing. This makes the gate harder to route when the layout is already occupied with other gates. Hence, these gates are prioritized to increase throughput. \ignore{We compute the Manhattan  lowerbound among (source, sink) tuples while iterating through them in Step 4.}

    \item \textbf{Route length:} Shorter paths have higher priority. For paths with equal number of (source, sink) pairs and the same Manhattan lowerbound, the shorter one is easier to accommodate with other routes. 
\end{enumerate}

Note that these rules are heuristics that can be replaced by better ones without changing the overall framework.

\subsubsection{Step 6 - Retry step} The rejected gates in Step 5 are given a second chance. We go through Steps 4 and 5 again for them. Since the number of unoccupied qubits in our layout is drastically lower at this stage, the W-BFS algorithm needs to find highly non-trivial paths (in Step 4) for routing the gate successfully in the current timestep (Fig \ref{fig:r2pqf_design_1}C).

\subsubsection{Step 7 - Update the set of pending gates} All the gates that are successfully routed in the current timestep are discarded from the $inj$ buffer/OS (based on their procurement source). At the same time, if any other PQF measurement fails during circuit execution, we add the corrective Gridsynth sequences into the $inj$ buffer.

Steps 1-7 are repeated until the $inj$ buffer is empty, i.e. no more Gridsynth gates to route.  

\subsection{Case 3: Post \protocol}\label{subsec:post_protocol}


\begin{figure}
    \centering
    \includegraphics[width=\linewidth]{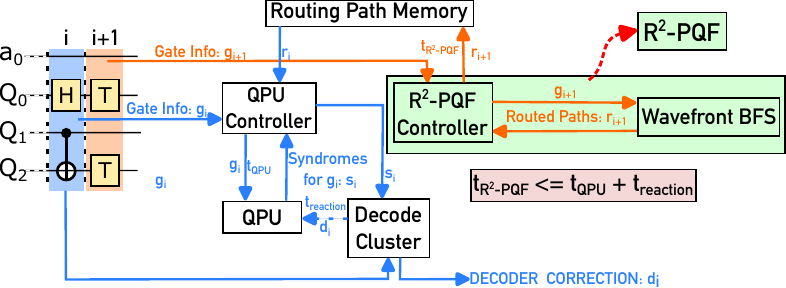}
    \caption{System level integration of \protocol. While gates in step ``\emph{i}'' are being executed (blue arrows), \protocol computes the routing paths for gates in step ``\emph{i+1}'' and stores them in Routing Path Memory (orange arrows). The execution time is $t_{QPU}/t_{QPU} + t_{reaction}$ (depending on whether we are implementing Clifford or non-Clifford gates). The decoder's processing overhead is not accounted since SOTA streaming/window decoders perform computation parallelly with syndrome generation on QPUs. The Decoder overhead reduces to classical communication costs associated with $T$ gate decoding ($t_{reaction}$) when relevant. We must complete executing \protocol for step ``\emph{i+1}'' in the same time that step ``\emph{i}'' executes.}
    \label{fig:r2_pqf_system_integration}
\end{figure}

Once we complete the execution of \protocol, we revert to using the routing paths in OS. However, the order of gates in OS might now be suboptimal since some of OS's originally concurrent gates may be routed in different time steps. We can insert gates from ``future'' time-steps - i.e. gates that were scheduled later in OS but can now be executed earlier since any potential conflicting gates in the earlier time-step have already been executed. We don't recompile these gates. If all the data dependencies of a ``future'' gate have been resolved in the current time-step, and it doesn't intersect with any existing route, then it can be routed immediately.

Figure \ref{fig:r2_pqf_system_integration} shows how \protocol integrates with the rest of the control stack. \revision{We mentioned earlier that \protocol computes the routing paths of gates at time-step $i+1$ while gates in step $i$ are being executed. This is valid for all time-steps except the first one post PQF failure. In this case, we stall circuit execution for one quantum clock cycle till \protocol completes path determination. The increase in schedule depth due to this stall is negligible -- on average $0.56\%$ and up to $1.07\%$ across all the circuits evaluated in Section \ref{sec:evaluation}. We have accounted for these stalls in all evaluations in Section \ref{sec:evaluation}.}
\section{Methodology}\label{sec:methodology}

\begin{figure}
    \centering
    \includegraphics[width=0.9\linewidth]{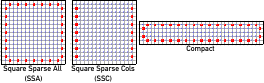}
    \caption{The three qubit layouts on which \protocol is evaluated. The blue circles represent surface code patches while the red squares represent magic state factories.}
    \label{fig:all_layouts}
\end{figure}

\begin{figure*}
    \centering
    \includegraphics[width=\linewidth]{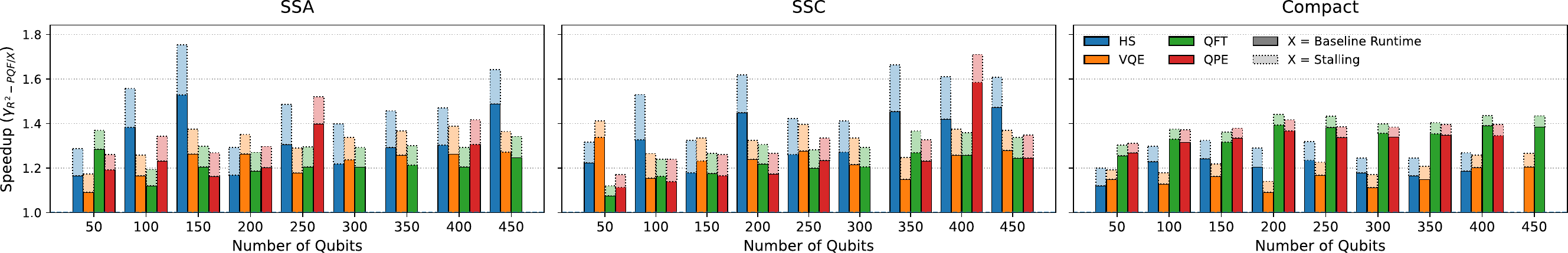}
    \caption{\revision{Speedup of \protocol over Stalling 
    ($\gamma_{R^2-PQF/Stalling}$) and a Baseline Runtime solution ($\gamma_{R^2-PQF/Baseline-Runtime}$)  across 50-450 qubits for four benchmarks across three layouts ($p_{succ}=90\%, \epsilon=10^{-13}$). The average and maximum values of $\gamma_{R^2-PQF/Stalling}$ for the three layouts are - $1.36\times$ and $1.76\times$ for SSA, $1.36\times$ and $1.71\times$ for SSC, and $1.31\times$ and $1.44\times$ for Compact. The average and maximum values of $\gamma_{R^2-PQF/Baseline-Runtime}$ for the three layouts are - $1.25\times$ and $1.53\times$ for SSA, $1.25\times$ and $1.59\times$ for SSC, and $1.26\times$ and $1.39\times$ for Compact. Note that these numbers include the additional cost for the first time-step post PQF measurement failure as discussed in the last paragraph of Section \ref{subsec:post_protocol}.}}
    \label{fig:variation_with_qubits}
\end{figure*}

\subsection{Benchmarks}\label{subsec:benchmarks} We evaluate \protocol for four benchmarks - Ising Model Hamiltonian Simulation (HS) from \cite{Supermarq} and VQE, QFT, and Phase Estimation (QPE) from \cite{mqt-bench}. Our benchmarks are evaluated for 10-450 qubit circuits. They are representative of (Early) Fault Tolerant applications and sweep across varying degrees of instruction parallelism.

\subsection{Layouts}\label{subsec:layouts}
We evaluate \protocol on three layouts - \textit{Square-Sparse-All} (SSA), \textit{Square-Sparse-Cols} (SSC), and \textit{Compact} (C). The layouts are shown in Figure \ref{fig:all_layouts}. In case of SSA, for an $n$-qubit program, the grid-dimensions are $(2\lceil \sqrt{n} \rceil + 3) \times (2\lceil \sqrt{n} \rceil + 3)$. This layout has a large ancilla routing space and is surrounded by Magic State Factories (MSFs) to ensure the computation is not $T$ limited and route congestion is spread in all directions. The SSC layout has the same dimensions, but MSFs only along the left and right edges. This increases contention for $T$ states and routing space, especially in highly parallel programs. Finally, the Compact layout has a dimension of $(2\lceil \frac{n}{2} \rceil + 1)\times 5$. It has a much lesser ancilla space but is surrounded by MSFs.

The three layouts represent low, medium, and high levels of ancilla space contention, respectively, forcing increasing serialization in the schedule. However, the Compact layout also has a shorter distance to MSFs, which results in shorter routing paths, lower likelihood of intersections,  and increased parallelism. \ignore{The Compact layout shows interesting trends in schedule length based on the dominating factor.}

\ignore{SSA/SSC are inspired by Litinski's fast block layout \cite{litinski2019game} which was originally designed for two-tile multi-qubit patches. We tweak it to make it compatible with single patch surface codes. The original fast block layout does not specify the placement of magic state factories. Hence, we assume factories distributed around the periphery and along the boundary columns of the layout for SSA and SSC respectively. Our Compact layout is the same as Litinski's compact layout with magic state factories around the periphery \cite{litinski2019game}.}

\subsection{Synthesis Parameters}\label{subsec:synthesis_params}

To evaluate the achievable clock frequency of the router on dedicated hardware, we implement \protocol in TSMC 16nm FinFET technology. The input RTL was generated from C using AMD's Vitis HLS and synthesized with Synopsys Design Compiler targeting the TSMC N16ADFP standard cell library. The design achieves timing closure at 2.0GHz under nominal operating conditions (0.8V, 25°C). The same RTL was also synthesized onto a Xilinx VU47P FPGA using Vivado, achieving 250 MHz. 

\subsection{Evaluation Metrics}\label{subsec:metric}

\begin{enumerate}
    \item \emph{Speedup} ($\gamma_{A/B}$): This metric quantifies the speedup obtained by protocol ``A'' over protocol ``B'' as the ratio of the schedule length of a circuit compiled using protocol ``B'' ($l_B$) to that of the schedule length of the same circuit compiled using protocol ``A'' ($l_A$).

    \begin{equation}
        \gamma_{A/B} = \frac{l_B}{l_A}
    \end{equation}

    \item \emph{Per-Timestep Latency} (PTL): This metric quantifies the router's classical time overhead at runtime. It is defined as the product of classical clock cycles per timestep (obtained from simulation) and the clock-time achieved through synthesis.
\end{enumerate}

f\section{Evaluation}\label{sec:evaluation}

\begin{figure}
    \centering
    \includegraphics[width=\linewidth]{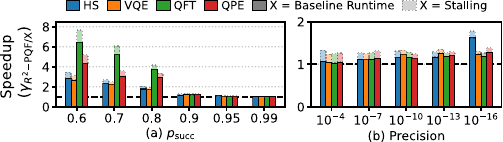}
    \caption{\revision{(a) Variation of Speedup across $p_{succ}$ ($\epsilon = 10^{-13}$). The average and maximum speedups of \protocol over Stalling ($\gamma_{R^2-PQF/SS}$) are $2.48\times$ and $7.67\times$ respectively. The average and maximum speedups of \protocol over the Baseline Runtime protocol are $2.18\times$ and $6.42\times$ respectively. At lower values of $p_{succ}$, the likelihood of PQF failure increases leading to a higher frequency of insertion of Gridsynth sequences. Hence, the effectiveness of \protocol increases. (b) Variation of Speedup across $\epsilon$ ($p_{succ} = 0.9$). The average and maximum speedups of \protocol over Stalling ($\gamma_{R^2-PQF/SS}$) are $1.32\times$ and $1.78\times$ respectively. The average and maximum speedups of \protocol over the Baseline Runtime protocol are $1.18\times$ and $1.62\times$ respectively. All data has been computed for the SSA layout and 200 qubits.}}\label{fig:variation_with_psucc_and_prec}
\end{figure}

\subsection{Comparison with Stalling Schedule and Baseline Runtime Compilation Schedule (BRCS)}\label{subsec:stalling_comparison}

\revision{The Stalling Schedule baseline has been described in Section \ref{subsec:stalling_sched}. In addition to SS, we also compare \protocol against a Baseline Runtime Compilation Schedule (BRCS) where we retain the paths in OS and allow parallel execution of non-intersecting runtime-inserted $T$ gates. The parallel execution would result in  BRCS having a higher throughput than SS and subsequently a shorter runtime. The paths for these $T$ gates are computed at compile-time assuming an empty grid and stored in memory. Any such path would be unaware of the route of a parallelly executing gate from OS at runtime. Hence, the likelihood of conflict and subsequent serialization is higher than in the case of \protocol. Comparison against BRCS quantifies the speedup due to \protocol's unique design decisions. In contrast, the SS baseline quantifies speedup due to both \protocol's design as well as the process of runtime recompilation itself. Note that unlike SS, which does not need any runtime logic, BRCS needs some lightweight logic to check for potential path intersections.}

\subsubsection{Variation of Speedup with number of qubits}\label{subsubsec:variation_with_qubits}
\revision{Figure \ref{fig:variation_with_qubits} shows the variation of $\gamma_{R^2-PQF/X}$ ($X = $ BRCS/SS) for circuits ranging from 50 to 450 qubits for all three layouts.} We observe fairly consistent performance across qubit counts for all benchmarks. While there is some stochasticity in the trends owing to the probabilistic nature of PQF, we observe that benchmarks with a large number of parallel $R_z$ gates - HS and VQE, generally perform better on unconstrained layouts - SSA and SSC. The SS serializes the program whenever an $R_z$ gate fails, turning highly parallel workloads serial. \protocol maintains program parallelism, effectively increasing the $\gamma_{R^2-PQF/SS}$ for highly parallel workloads. Note that these effects are dampened in the Compact layout since parallel routing is hindered by the lack of adequate ancilla space. Thus, $\gamma_{R^2-PQF/SS}$ depends both on the circuit structure and the layout. \revision{We also observe that $\gamma_{R^2-PQF/BRCS}$ is consistently lower than $\gamma_{R^2-PQF/SS}$ for all workloads highlighting the separate contributions of \protocol's design and the process of runtime-recompilation itself. 
The average and maximum values of $\gamma_{R^2-PQF/SS}$ and $\gamma_{R^2-PQF/BRCS}$ are reported in the caption of Figure \ref{fig:variation_with_qubits}. The missing data points in Figure \ref{fig:variation_with_qubits} correspond to jobs that timed out on our compute cluster.}

\revision{Circuits with $\leq 100$ qubits were executed 5 times to gauge \protocol's variance. Across all workloads, we observed an average and maximum variance of $5.28\%$ and $11.00\%$ for $\gamma_{R^2-PQF/SS}$ and $5.35\%$ and $11.08\%$ for $\gamma_{R^2-PQF/BRCS}$. Circuits with more qubits were excluded from this computation due to high computational overhead of executing them. We also executed \protocol for the $\leq 100$ qubit workloads with rotating qubits instead of fixed ones and compared the values of speedup against the fixed-qubit counterparts. On average, the rotating qubit version of \protocol results in the $\gamma_{R^2-PQF/SS}$ ($\gamma_{R^2-PQF/BRCS}$) increasing on average by $3.6\%$ ($3.8\%$) and up to $9.1\%$ ($8.7\%$).}

\subsubsection{Variation of Speedup w.r.t. $p_{succ}$}\label{subsubsec:variation_with_psucc}

\revision{Figure \ref{fig:variation_with_psucc_and_prec}a shows the variation of $\gamma_{R^2-PQF/X} (X = SS/BRCS)$ with $p_{succ}$. As expected, we see a drop in $\gamma_{R^2-PQF/X}$ as $p_{succ}$ increases. For low values of $p_{succ}$, the chances of PQF failure increase, which increases the number of Gridsynth sequences inserted in the circuit. Since R2S enables a higher degree of parallelism than SS or BRCS, the relative speedup increases with decreasing $p_{succ}$. The average and maximum $\gamma_{R^2-PQF/X}$ are mentioned in the caption of Figure \ref{fig:variation_with_psucc_and_prec}.}

\subsubsection{Variation of Speedup with precision ($\epsilon$)}\label{subsubsec:variation_with_precision}

\revision{Figure \ref{fig:variation_with_psucc_and_prec}b shows the variation of $\gamma_{R^2-PQF/X} (X = SS/BRCS)$ with $\epsilon$. For a fixed layout and benchmark, $\gamma_{R^2-PQF/X}$ remains largely constant w.r.t. $\epsilon$. While a lower value of $\epsilon$ leads to longer routing schedules, the factor of increment is the same across SS/BRCS/R2S. This factor cancels out from the numerator and denominator in our definition of $\gamma$, which results in a generally uniform trend across $\epsilon$.}

\subsection{Comparison with Global Recompilation Schedule}\label{subsec:recompilation_comparison}

\subsubsection{Variation of $\gamma_{Global\_Recompilation/R^2-PQF}$ with qubits}\label{subsubsec:overshoot_with_qubits}

\begin{figure}
    \centering
    \includegraphics[width=0.8\linewidth]{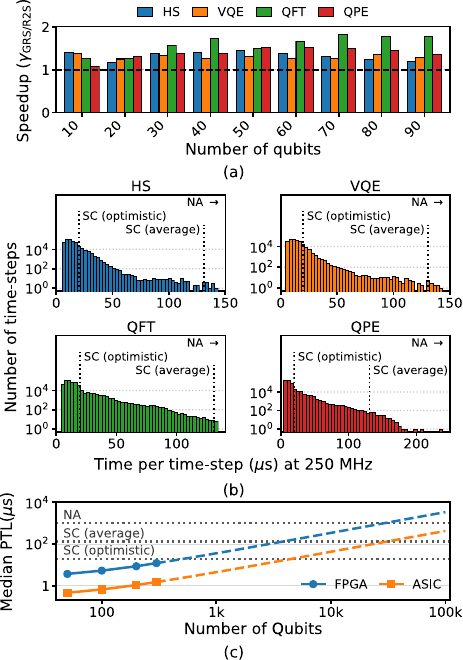}
    \caption{\revision{(a) Variation of $\gamma_{Global\_Recompilation/R^2-PQF}$ with qubits for SSA layout, $\epsilon = 10^{-13}, p_{succ} = 0.9$ The average and max values of $\gamma_{Global\_Recompilation/R^2-PQF}$ for the evaluated benchmarks are  $1.34 \times$ and $1.73 \times$ respectively (b) Histogram showing the distribution of PTL on an FPGA for 300-qubit circuits of all four bencharks (c) Scaling of Median PTL with circuit size and comparison against time-budget afforded by different technolgoies.}}
    \label{fig:overshoot_fig}
\end{figure}

Figure \ref{fig:overshoot_fig}a compares the length of GRS (Global Recompilation) with that of R2S (\protocol). Note that GRS can only be evaluated for only a small number of qubits 
since it has a much larger computational overhead than both SS and R2S and times out for much smaller qubit numbers. The average and max values of $\gamma_{Global\_Recompilation/R^2-PQF}$ for the evaluated benchmarks are  $1.34 \times$ and $1.73 \times$ respectively.

\subsection{Comparison against Pauli Product Rotations}\label{subsec:ppr_comparison}

\begin{figure}[h!]
    \centering
    \includegraphics[width=0.8\linewidth]{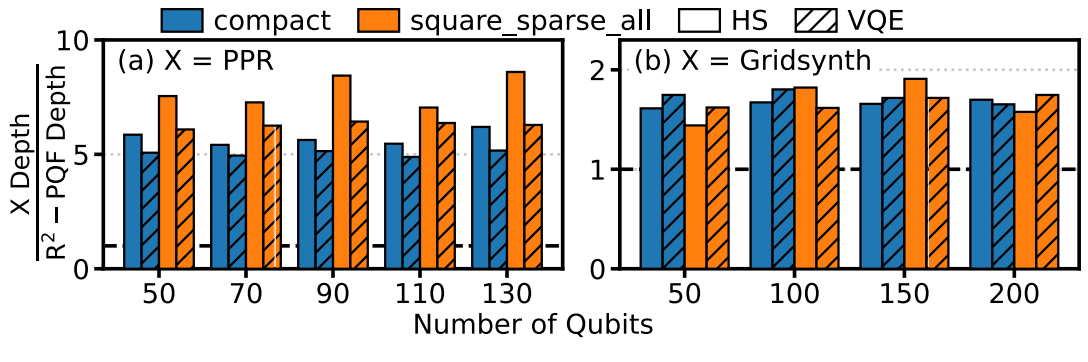}
    \caption{\revision{(a)PPR circuits are on average $6.1\times$ and up to $8.6\times$ deeper than \protocol compiled Clifford+$T$ circuits. (b) Gridsynth circuits are on average $1.68\times$ and upto $1.91\times$ deeper than \protocol compiled PQF circuits.}}
    \label{fig:ppr_gridsynth_depth_comparison}
\end{figure}

As mentioned in Section \ref{subsec:surface_codes_synthesis}, and reported in Ref \cite{wang2024optimizing}, a PPR basis often results in much longer sequences than Clifford+T for Gridsynth synthesized circuits. Since a PQF Clifford+$T$ synthesis has fewer gates than its Gridsynth counterpart, we expect PQF to outperform PPRs in terms of circuit depth (and hence duration) as well. Figure \ref{fig:ppr_gridsynth_depth_comparison}a compares the depth of PPR circuits \protocol-compiled Clifford+$T$ circuits on the same layouts. We use the PPR compilation workflow given in Ref \cite{litinski2019game}. We observe that PPR circuits are on average $6.1\times$ and up to $8.6\times$ deeper than \protocol circuits.

\subsection{\revision{Comparison against Gridsynth}}\label{subsec:gridsynth_comp}

\revision{Figure \ref{fig:ppr_gridsynth_depth_comparison}b shows the speedup of PQF circuits vs Gridsynth circuits. They are compiled using \protocol and DASCOT-G, respectively. Since Gridsynth-synthesised sequences are much deeper than PQF-synthesized sequences, we expect the Gridsynth schedule to be deeper. For the benchmarks considered, PQF circuits are on average $1.68\times$ and up to $1.91\times$ faster than Gridsynth circuits (Keeping layout, physical qubits and Magic State Factories the same).}

\subsection{Hardware Latency}\label{subsec:hardware_latency}

\begin{table}[t]
\centering
\scriptsize
\setlength{\tabcolsep}{1.5pt}
\resizebox{\columnwidth}{!}{%
\begin{tabular}{|l|rr|rr|rr|cc|cc|}
\hline
 & \multicolumn{2}{c|}{\textbf{Cyc/step}} & \multicolumn{2}{c|}{\textbf{FPGA} ($\mu$s)} & \multicolumn{2}{c|}{\textbf{ASIC} ($\mu$s)} & \multicolumn{2}{c|}{\textbf{FPGA \%}} & \multicolumn{2}{c|}{\textbf{ASIC \%}} \\
\textbf{Benchmark} & \textbf{Med.} & \textbf{Max} & \textbf{Med.} & \textbf{Max} & \textbf{Med.} & \textbf{Max} & \textbf{$<t_{r}$} & \textbf{$<t_{Q}$} & \textbf{$<t_{r}$} & \textbf{$<t_{Q}$} \\
\hline
QFT\_50  & 0.9k & 10k & 3.58 & 40.6 & 0.45 & 5.08 & 93.4 & 99.3 & 100.0 & 100.0 \\
QFT\_100 & 1.3k & 13k & 5.28 & 50.2 & 0.66 & 6.27 & 87.1 & 98.0 & 100.0 & 100.0 \\
QFT\_200 & 2.2k & 29k & 8.64 & 116  & 1.08 & 14.6 & 63.0 & 94.0 & 99.9  & 100.0 \\
QFT\_300 & 2.7k & 34k & 11.0 & 134  & 1.37 & 16.8 & 42.6 & 90.3 & 99.7  & 100.0 \\
\hline
QPE\_50  & 0.8k & 12k & 3.19 & 48.0 & 0.40 & 6.00 & 94.4 & 99.4 & 100.0 & 100.0 \\
QPE\_100 & 1.3k & 11k & 5.27 & 45.2 & 0.66 & 5.65 & 91.7 & 99.0 & 100.0 & 100.0 \\
QPE\_200 & 2.1k & 35k & 8.38 & 139  & 1.05 & 17.4 & 66.9 & 94.8 & 99.9  & 100.0 \\
QPE\_300 & 2.7k & 60k & 11.0 & 238  & 1.37 & 29.8 & 45.9 & 89.7 & 99.5  & 99.9  \\
\hline
VQE\_50  & 0.9k & 15k & 3.67 & 60.5 & 0.46 & 7.56 & 97.7 & 99.7 & 100.0 & 100.0 \\
VQE\_100 & 1.3k & 18k & 5.35 & 73.4 & 0.67 & 9.18 & 93.5 & 99.8 & 100.0 & 100.0 \\
VQE\_200 & 2.2k & 66k & 8.68 & 263  & 1.09 & 32.8 & 60.3 & 95.8 & 99.8  & 99.9  \\
VQE\_300 & 3.0k & 36k & 12.1 & 143  & 1.51 & 17.9 & 35.2 & 89.6 & 99.9  & 100.0 \\
\hline
HS\_50   & 0.9k & 15k & 3.67 & 60.5 & 0.46 & 7.56 & 97.6 & 99.8 & 100.0 & 100.0 \\
HS\_100  & 1.3k & 18k & 5.28 & 73.4 & 0.66 & 9.18 & 92.2 & 99.8 & 100.0 & 100.0 \\
HS\_200  & 2.1k & 66k & 8.53 & 263  & 1.07 & 32.8 & 61.8 & 93.7 & 99.9  & 99.9  \\
HS\_300  & 3.0k & 36k & 12.1 & 143  & 1.51 & 17.9 & 35.2 & 90.1 & 99.9  & 100.0 \\
\hline
\end{tabular}%
}
\vspace{6pt}
\caption{Per-timestep latency and timing budget compliance.
 FPGA: 250\,MHz on Xilinx VU47P. ASIC: 2.0\,GHz on TSMC 16nm.
$t_{r}=t_{react}=10\,\mu s$ and $t_{Q}\equiv t_{QPU}{+}t_{r}=20\,\mu s$,
at $d=10$.}
\label{tab:latency}

\end{table}

\revision{As mentioned in Section \ref{subsubsec:runtime_path_determination}, the time budget for \protocol depends on the operation, code distance, reaction time and underlying technology. However, we use a conservative $20\mu s$ value for $T$ sequences. At 2.0GHz, the median PTL for all four benchmarks is well within the \revision{$20\mu s$} latency budget for all qubit sizes. Across all benchmarks, the PTL exceeds \revision{$20\mu s$} for at most \revision{$0.1\%$} of the time-steps (Table \ref{tab:latency} - ASIC$\% < t_Q$ )}

\revision{In most cases, this won't be an issue since the time-budget in practice might be much larger due to a larger code distance ($d>10$), longer reaction time ($t_{reaction} > 10\mu s$) or slower technology (Neutral atoms or trapped ions). However, in the rare case where the time-budget equals the conservative estimate, we stall the program till \protocol completes execution. This is a stall of at most 1 time-step which results in the schedule depth increasing at most by $0.1\%$. For reference, the range of R2S depth for the benchmarks in Figure \ref{fig:variation_with_qubits} is [12146, 306311] time-steps and a $0.1\%$ increase corresponds to at most 307 clock cycles, which for a $20\mu s$ clock cycle corresponds to at most 0.006 seconds.} 

\revision{On the FPGA at 250 MHz, the PTL of most time-steps is also within the $20\mu s$ budget, however we have more outliers comapred to the much faster ASIC. Table \ref{tab:latency} - FPGA$\% < t_Q$ shows that across all workloads, $89.6\%$ to $98\%$ of the time-steps are within the conservative $20\mu s$ budget. As workload sizes increase, the number of time-steps that exceed $20 \mu s$ also increase. Fig \ref{fig:overshoot_fig}B shows the histogram of latencies of different time-steps for 300 qubit instances of all four benchmarks. We also plot the lines corresponding to the time budget for Superconducting (SC) (optimistic) - $20\mu s$, and Superconducting (SC) (average) - $131 \mu s$. The Neutral Atom (NA) budget is 1 mili-second and is much farther to the right as shown by the arrow in all subplots of Fig \ref{fig:overshoot_fig}B.} 

\revision{Figure \ref{fig:overshoot_fig}C shows the scaling of Median PTL of \protocol with the number of logical qubits and compare it with the SC (optimistic), SC (average) and NA time-budgets. The FPGA and ASIC trends have been extrapolated from the data in Table \ref{tab:latency}. An FPGA implementation is real-time feasible upto $\sim 500$, $3.8$k and $30$k logical qubits for the three time-budgets. The corresponding values for a ASIC implementation are $\sim 4.5$k, $31$k, and $240$k -- all circuits well within the quantum utility range. Since there is a trade-off between \protocol's capabilities and its time overhead, one can possibly design a protocol with more complex logic and better performance if the time-budget far exceeds \protocol's time overhead. For example, for a 100 qubit circuit on Neutral atoms, the time-budget is $100-1000\times$ greater than the time cost of \protocol.}

\subsection{\revision{Hardware Utilization}}\label{subsec:hardware_utils}

\revision{Table ~\ref{tab:hardware} summarizes the hardware characteristics of each synthesized design with a qubit count of 100, and grid size of $70 \times 64$. Table~\ref{tab:scaling} reports FPGA utilization for configurations supporting 50
to 300 logical qubits, each with its corresponding SSA grid. Each configuration achieves timing closure at 250 MHz, and logic grows far more slowly than the
qubit count: a 6$\times$ increase in supported qubits costs only 29\% more flip-flops and 28\% more LUTs. BRAM is the one resource that scales with qubit count, rising from 12\% to 89\%. The fifth column isolates grid size from qubit count. The over-provisioned $70{\times}64$ configuration of Table~\ref{tab:hardware} supports the same 100
qubits as column two using identical BRAM, but 62\% more LUTs, confirming that memory tracks qubit count while logic tracks grid dimension. Across the board the absolute footprint is modest, all fitting on less than one full FPGA.} 

\begin{table}[t]
\centering
\footnotesize
\setlength{\tabcolsep}{3pt}
\begin{tabular}{|l|l|l|}
\hline
\textbf{Metric} & \textbf{FPGA (VU47P)} & \textbf{ASIC (TSMC 16nm)} \\
\hline
Synthesis tool & Vitis HLS + Vivado & Synopsys DC V-2023 \\
Target frequency & 250 MHz & 2.0 GHz (TT) \\
Grid / qubits & 70$\times$64 / 100 & 70$\times$64 / 100 \\
Flip-flops / Seq. Cells & 395,913 flip-flops & 176,057 seq. cells \\
Combinational cells & --- & 428,740 \\
LUT utilization & 389,815 (30\%) & --- \\
BRAM utilization & 445.5 (22\%) & --- \\
DSP utilization & 12 & --- \\
Area & --- & 0.223 mm$^2$ \\
Dynamic power & 6.69 W (vectorless) & 1.71 W (vectorless) \\
Static / leakage power & 3.42 W (whole device) & 3.61 mW \\
\hline
\end{tabular}
\vspace{6pt}
\caption{\revision{Hardware Characteristics of each Synthesized Design}}
\label{tab:hardware}
\end{table}

\begin{table}[t]
\centering
\footnotesize
\setlength{\tabcolsep}{3pt}
\begin{tabular}{|l|r|r|r|r|r|}
\hline
 & \multicolumn{4}{c|}{\textbf{Supported qubits}} & \\
\textbf{Metric} & \textbf{50} & \textbf{100} & \textbf{200} & \textbf{300} & \textbf{100$^{*}$} \\
\hline
Grid             & 19$\times$19 & 23$\times$23 & 33$\times$33 & 39$\times$39 & 70$\times$64 \\
Flip-flops       & 235,891 & 251,442 & 282,568 & 304,352 & 395,913 \\
LUT utilization  & 211,694 & 240,741 & 254,752 & 271,055 & 389,815 \\
                 & (16\%) & (18\%) & (20\%) & (21\%) & (30\%) \\
BRAM utilization & 243.5 & 445.5 & 977.5 & 1801.5 & 445.5 \\
                 & (12\%) & (22\%) & (48\%) & (89\%) & (22\%) \\
DSP utilization  & 12 & 12 & 12 & 12 & 12 \\
\hline
\end{tabular}
\vspace{6pt}

\caption{\revision{FPGA resource scaling with supported qubit count. $^{*}$The over-provisioned configuration
reported in Table~\ref{tab:hardware}: same qubit count as column two with a larger grid.}}
\label{tab:scaling}
\end{table}

\ignore{The subsequent memory error is negligible since stalling instances are very rare ($< 0.5\%$ of all time-steps). Further, we never stall for more than a couple of quantum clock cycle - i.e. no prolonged stalls as is the case with GRS. Each stall introduces an error of $\simeq 10^{\frac{-(d+1)}{2}}$ ($d=$ code distance), which for practical code distances ($d \geq$ 13) is $\leq 10^{-7}$. The longest program in our evaluation is $QFT\_300$ which takes 150692 timesteps when routed using \protocol. The total error introduced from stalling is $0.007\%$ for $d=13$. It is lesser for larger code distances.}

\section{Discussion}\label{sec:discussion}

\subsection{\revision{W-BFS vs conventional BFS}}\label{subsec:w_bfs_vs_conv_bfs}
\revision{W-BFS is chosen over conventional BFS (Conv-BFS) due to its parallelizability, which results in faster path computation. Both BFS variants produce paths of the same length by construction. W-BFS's speed comes at a higher hardware cost. More state bits and registers are needed to process signal propagation patterns in W-BFS compared to Conv-BFS.}  

\revision{To compare the two, we randomly choose two points in a square grid and compute the path length in both cases. We repeat this for $10^5$ trials and report the number of instances when the path lengths differ ($N_{mismatch}$ -- always expected to be 0, the ratio of their mean clock cycle times ($R_1 = \frac{T_{mean} (Conv-BFS)}{T_{mean} (W-BFS)}$), maximum clock cycle times ($R_2$), state-bit costs ($R_3$) and logic element costs ($R_4$). We report the data in Table \ref{tab:w_bfs_conv_bfs_comparison}. As expected, $N_{mismatch} = 0, R_1 > 1, R_2 > 1, R_3 < 1, $and $R_4 < 1$ for all cases.}

\begin{table}[h!]
    \centering
    \begin{tabular}{|c|c|c|c|c|c|}
    \hline
    $L_{grid}$ & $N_{mismatch}$ & $R_1$ & $R_2$ & $R_3$ & $R_4$ \\
    \hline
    27 & 0 & 26.50 & 29.29 & 0.62 & 0.049\\
    \hline
    41 & 0 & 46.63 & 47.69 & 0.58 & 0.021 \\
    \hline
    55 & 0 & 68.16 & 67.10& 0.55 & 0.019\\ 
    \hline
    \end{tabular}
    \vspace{6pt}
    \caption{\revision{Solution quality ($N_{mismatch}$), latency ($R_1$ and $R_2$), and hardware cost ($R_3$ and $R_4$) comparisons between W-BFS and Conventional BFS. See Section \ref{subsec:w_bfs_vs_conv_bfs} for notation.}}
    \label{tab:w_bfs_conv_bfs_comparison}
\end{table}

\subsection{\revision{Ablation Studies}}\label{subsec:ablation}

\subsubsection{\revision{Unidirectional vs Bi-directional W-BFS}}\label{subsubsec:uni_v_bi}
\revision{All variants of BFS, by construction, produce same length paths. Hence, replacing bidirectional W-BFS with its unidirectional variant produces insignificant change in schedule depth. Bidirectional BFS converges $\sim 2\times$ faster; however, both variants have the same setup and backtracking costs, which reduces the net gain. For square grids of side length $L = 27, 41, 55$,  bi-directional W-BFS is on average $1.17\times, 1.20\times, 1.22\times$ faster than unidirectional W-BFS, respectively.}

\subsubsection{\revision{Arbitration rules, Retry stage, ASAP insertion}}\label{subsubsec:remaining_ablation}

\revision{The remaining stages impact the depth of R2S. We compute the contribution of each stage using Shapley values - a common method used to gauge the impact of each component when they aren't strictly orthogonal \cite{shapley1953value, lundberg2017unified}. We report the speedup contribution of base-\protocol (where all additional heuristics are switched off) and the differential contribution of each heuristic for VQE and HS for the SSA and Compact layouts averaged across qubits ranging from $20$ to $400$ for $p_{succ} = 0.9$ and $\epsilon = 10^{-13}$ in Table \ref{tab:ablation_data}. Note that there are cases when a stage has a negative contribution, i.e. $depth(R2S)$ reduces when the component is eliminated. This is expected since the stages are heuristics that may be detrimental in some cases.} 
\revision{However, no stage contributes negatively across all instances, and they all have a net positive contribution. For SSA, the retry stage has the largest contribution for both HS and VQE. This is expected since there is substantial routing space in the layout and a second attempt likely results in the compiler finding a valid path. The retry stage has $\sim 0$ (rather a slight negative) contribution in the Compact layout. This is due to scarce routing space that makes a retry stage redundant. Hence, the first and second arbitration rules contribute most of the speedup since the routes must be computed optimally in the first attempt. These insights can be used to make layout-specific optimizations to the \protocol pipeline and further reduce its computational overhead.} 

\begin{table}[ht!]
    \centering
    \begin{tabular}{|c|c|c|c|c|}
    \hline
         Component & HS + SSA & QFT + SSA & HS + C & QFT + C \\
         \hline
         base-\protocol & 60.2 & 57.4 & 21.6 & 11.9 \\
         \hline
         Arb-rule 1 & 7.6 & 9.6 & 39.3 & 44.9 \\
         \hline
         Arb-rule 2 & -8.6 & 0.9 & 37.1 & 43.6 \\
         \hline
         Arb-rule 3 & 7.0 & -0.2 & -2.6 & -2.4 \\
         \hline
         Retry & 31.1 & 32.2 & -0.1 & -0.4 \\
         \hline
         ASAP & 2.7 & 0.2 & 4.6 & 2.4 \\
         \hline
    \end{tabular}
    \vspace{6pt}
    \caption{\revision{Differential contribution of various components of the \protocol pipeline as a percentage ($\%$) of the overall speedup. See section \ref{subsubsec:remaining_ablation} for a detailed discussion.}}
    \label{tab:ablation_data}
\end{table}

\subsection{Related Work}\label{subsec:related_work}

\subsubsection{\textbf{Surface Code Compilation}}For Surface codes, various compilers have been proposed to handle qubit mapping and routing. 
DASCOT \cite{molavi2025dependency} optimizes the circuit mapping and routing using simulated annealing. Early surface code compilers either didn't optimize for execution time (LSC \cite{qec_compiler1}) or optimize routing for a trivial mapping on a standard layout (EDPC \cite{qec_compiler2}). All these works handle the compilation of static circuits only - i.e., circuits that are the same at compile time and runtime. In contrast, \protocol allows for the compilation of PQF enabled dynamic circuits while optimizing circuit execution time.

\ignore{\subsubsection{\textbf{Dynamic Circuit Compilation}}While \protocol has been designed to handle dynamism from PQF synthesis, there continue to exist other sources of dynamism, especially in error corrected circuits. As mentioned in Section \ref{sec:introduction}, $T$ gate consumption circuits often require a dynamically inserted $S$ gate. While a single $S$ gate is much shorter than a long Gridsynth sequence, a $T$ consumption fails much more often ($50\%$) than a PQF synthesis does ($\sim1-p_{succ}$). Thus, recompiling might lead to significant benefits in circuit runtime here. \protocol can be easily repurposed for this case too. Dynamic circuits have shown to reduce the depth of key quantum subroutines like GHZ preparation and gate teleportation and enabled their experimental realization on NISQ hardware \cite{baumer2024efficient}. This dynamism is in the construction of the circuit, orthogonal to the one enforced by error corrected execution of gates. Despite such promising applications and a pressing need, managing dynamism in circuits remains understudied. We encourage more future work in this direction - especially in generalizing runtime recompilation to other sources of dynamism too.}

\section{Conclusion}\label{sec:conclusion}

Probabilistic Quantum Circuits with Fallback (PQF) synthesizing $R_z$ gates with $\sim 3\times$ fewer $T$ gates compared to Gridsynth. However, it is non-deterministic and the synthesis may fail at runtime warranting the insertion of corrective gates. Routing these gates in conjunction with the circuit's pre-existing gates is needed to minimize circuit duration. However, this warrants the development of compilation tools that are fast enough to operate at runtime. \protocol does exactly this - it dynamically routes gates layer-by-layer using a Wavefront-BFS algorithm at runtime. \protocol speeds up circuit execution by $1.34\times$ on average and up to $1.76\times$ while reliably completing execution within the strict time budget enforced by the quantum clock cycle.


\bibliographystyle{IEEEtranS}
\bibliography{refs}

\end{document}